\documentclass[11pt,a4paper]{article}
\pdfoutput=1
\usepackage{jheppub}

\DeclareMathOperator{\Tr}{Tr}

\newcommand{\ri}{\mathrm{i}}

\newcommand{\qu}{\frac{1}{4}}

\newcommand{\Si}{\Sigma}
\renewcommand{\b}[1]{\overline{#1}}
\newcommand{\del}{\partial}

\newcommand{\bra}{\langle}
\newcommand{\ket}{\rangle}
\newcommand{\la}{\lambda}

\newcommand{\ka}{\kappa}

\newcommand{\bt}{\beta}
\newcommand{\ga}{\gamma}

\newcommand{\al}{\alpha}

\newcommand{\rt}[1]{\sqrt{#1}}

\begin{document}

\title{Eigenvalue instantons in the spectral form factor of random matrix model}

\author{Kazumi Okuyama}

\affiliation{Department of Physics, 
Shinshu University,\\
3-1-1 Asahi, Matsumoto, Nagano 390-8621, Japan}

\emailAdd{kazumi@azusa.shinshu-u.ac.jp}

\abstract{
We study the late time plateau behavior of the spectral form factor in
the Gaussian Unitary Ensemble (GUE) random matrix model.
The time derivative of the spectral form factor
in the plateau regime is not strictly zero, but non-zero due to
a non-perturbative correction in the $1/N$ expansion.
We argue that such a non-perturbative correction
comes from the eigenvalue instanton of random matrix model
and we explicitly compute the instanton correction as a function of time.
}

\maketitle

\section{Introduction}
The spectral form factor (SFF) is first
introduced in
\cite{jost} as a Fourier transform of the two-level correlation
function of energy eigenvalues. It 
is a very basic and important quantity in the study of quantum chaos;
it is widely used in many areas of physics ranging from nuclear physics to
condensed matter physics, to name a few, and even the application to
quantum gravity and black holes is discussed in the literature \cite{Cotler:2016fpe}.
At late times, SFF of a chaotic system
exhibits a universal behavior, so-called ramp and plateau, and it
is described by a random matrix theory of a certain ensemble
of matrices depending on the symmetry of the system.

As an application to quantum gravity,
the SFF in the Sachdev-Ye-Kitaev (SYK) model \cite{KitaevTalks,Sachdev,Maldacena:2016hyu}
has been intensively studied
as a useful diagnostic of the chaotic behavior of the
dual black hole in two dimensions \cite{Cotler:2016fpe}.
It is observed numerically that the SFF of the SYK model
indeed exhibits the structure of ramp and plateau as expected for a chaotic system.
In the recent paper \cite{Saad:2018bqo},
the structure of ramp was studied from the viewpoint of the collective field theory
in terms of the variables $G,\Si$, and the possible interpretation of the bulk geometry was
discussed. However, the structure of plateau in the SYK model is not yet fully understood.
It is speculated that the plateau in the SYK model
is closely related to the non-perturbative effects in the large $N$ expansion \cite{Cotler:2016fpe,Saad:2018bqo}. 

Motivated by this conjecture in the SYK model,
in this paper we will consider a much simpler example
of the plateau in the Gaussian Unitary Ensemble (GUE) random matrix model. 
The SFF in the GUE random matrix model
was extensively studied in the literature and the
exact results in the large $N$ limit  as well as the exact results at finite $N$ 
are available (see e.g. \cite{hikami, Liu:2018hlr, Okuyama:2018yep}).
In the plateau regime,
the SFF is approximately constant in time
and its time derivative vanishes in the large $N$ limit.
However, at finite
$N$ it is non-zero  due to a non-perturbative effect in the $1/N$ expansion.

We will argue that such a non-perturbative correction comes from the eigenvalue
instanton in the matrix integral, where some of the eigenvalues are pulled out
from the cut and put on the non-trivial saddle points 
of the eigenvalue integral \cite{David:1990sk,David:1992za}.\footnote{
A collective field approach to the non-perturbative corrections to the SFF
was studied by Andreev and Altshuler in \cite{Altshuler}. We will not consider that approach in this
paper.}
Naively, one might think that there is no non-perturbative correction 
in the $1/N$ expansion of Gaussian matrix model.
However, we should recall that by the ``late time'' we mean the following scaling limit
\begin{equation}
\begin{aligned}
 N\to\infty,~t\to\infty,\quad\text{with}~~\tau=\frac{t}{2N}~~\text{fixed}.
\end{aligned} 
\label{eq:tau}
\end{equation}
Since the SFF is defined by inserting the operator $\Tr e^{\pm\ri tH}=
\Tr e^{\pm 2\ri N\tau H}$
into the matrix integral, 
there appears a non-trivial saddle point in the scaling limit \eqref{eq:tau}
due to the $\mathcal{O}(N)$ potential exerted on the eigenvalues.
It turns out that the saddle points are located on the imaginary axis
since the exponent of $\Tr e^{\pm 2\ri N\tau H}$ is pure imaginary.

Interestingly, the computation of the non-perturbative correction
to the SFF in the plateau regime
is formally equivalent to
the similar computation in the 2d Yang-Mills theory on $S^2$
at weak coupling.
In fact, the one-point function of the operator $\Tr e^{\pm 2\ri N\tau H}$
in the GUE matrix model, whose exact form at finite $N$ 
is given by the Laguerre polynomial
\eqref{eq:Zexact}, is exactly the same as the 
expression appeared in the study of the partition function of
2d Yang-Mills theory on $S^2$ \cite{Gross:1994mr} under a certain identification
of the parameters.
This implies that, at least technically,
the ramp-plateau transition in the SFF
of the GUE random matrix model
is related to the Douglas-Kazakov phase transition of the 2d Yang-Mills theory on $S^2$
\cite{Douglas:1993iia}.

This paper is organized as follows.
In section \ref{sec:SFF}, after reviewing the exact result 
of the SFF in the GUE random matrix model,
we consider the non-perturbative correction to
the SFF using the formal relation to
the non-perturbative correction in the 2d Yang-Mills theory on $S^2$.
In section \ref{sec:inst}, we argue that the non-perturbative correction to the 
SFF comes from the eigenvalue instantons.
In section \ref{sec:1/N}, we compute the $1/N$ corrections around the instanton
configuration using the differential equation obeyed by
the Laguerre polynomials.
In section \ref{sec:beta}, we briefly comment on the case of the non-zero
inverse temperature $\bt\ne0$.
Finally, we conclude in section \ref{sec:conc}.

\section{Exact spectral form factor at $\bt=0$ \label{sec:SFF}}
Let us first review the exact form of the SFF in the GUE random matrix model
at finite $N$ obtained in \cite{Okuyama:2018yep}.
In the following, we will set the inverse temperature $\bt=0$ for simplicity.
We will comment on the $\bt\ne0$ case in section \ref{sec:beta}.

The SFF in the GUE random matrix model is defined by
\begin{equation}
\begin{aligned}
 g(\tau)=\Bigl\langle\Tr e^{\ri t H}\Tr e^{-\ri t H}\Bigr\rangle,
\end{aligned} 
\label{eq:gdef}
\end{equation}
where the expectation value is given by the Gaussian integral over
the $N\times N$ hermitian matrix $H$
\begin{equation}
\begin{aligned}
 \bigl\langle\cdots\bigr\rangle=
\frac{\int dH e^{-\frac{N}{2}\Tr H^2}(\cdots)}{\int dH e^{-\frac{N}{2}\Tr H^2}}.
\end{aligned} 
\end{equation}
We are interested in the late
time behavior of the SFF
in the scaling limit \eqref{eq:tau}, and we use $\tau=t/2N$ as
the time coordinate in what follows.
The SFF is naturally
decomposed into the disconnected part
$g_{\text{disc}}(\tau)$ and the connected part $g_{\text{conn}}(\tau)$ 
\begin{equation}
\begin{aligned}
  g(\tau)
=g_{\text{disc}}(\tau)+
g_{\text{conn}}(\tau).
\end{aligned} 
\end{equation}
The disconnected part is given by the product of one-point functions
\begin{equation}
\begin{aligned}
 g_{\text{disc}}(\tau)=\mathcal{Z}(\tau)^2,
\end{aligned} 
\label{eq:disc-z^2}
\end{equation}
with $\mathcal{Z}(\tau)$ being the one-point function
\begin{equation}
\begin{aligned}
 \mathcal{Z}(\tau)=\Bigl\langle\Tr e^{\ri t H}\Bigr\rangle=\Bigl\langle\Tr e^{-\ri t H}\Bigr\rangle.
\end{aligned} 
\label{eq:Zdef}
\end{equation}
Note that  $\mathcal{Z}(\tau)$ is an even function of $\tau$
and it does not dependent on the sign of $\tau$.
The exact form of $\mathcal{Z}(\tau)$ at finite $N$ is given by
\begin{equation}
\begin{aligned}
 \mathcal{Z}(\tau)=e^{-2N\tau^2}L_{N-1}^1(4N\tau^2),
\end{aligned} 
\label{eq:Zexact}
\end{equation}
where $L^\al_n(x)$ denotes the associated Laguerre polynomial.
One can also write down the exact form of the connected part
$g_{\text{conn}}(\tau)$ as a trace of some $N\times N$ matrix \cite{Okuyama:2018yep}.
It turns out that the time derivative of the connected part
$\del_\tau g_{\text{conn}}(\tau)$ has a simple expression \cite{Okuyama:2018yep}
\begin{equation}
\begin{aligned}
 \frac{\del g_{\text{conn}}(\tau)}{\del\tau}
=8N^2\tau e^{-4N\tau^2}\Bigl[L_{N-1}(4N\tau^2)L_{N-1}^1(4N\tau^2)-
L_{N}(4N\tau^2)L_{N-2}^1(4N\tau^2)\Bigr].
\end{aligned} 
\label{eq:gconn-exact}
\end{equation}
Moreover, using the property of Laguerre polynomials, one can show that
the connected part $g_{\text{conn}}(\tau)$ and 
the disconnected part $g_{\text{disc}}(\tau)$ are related by \cite{hikami}
\begin{equation}
\begin{aligned}
 g_{\text{disc}}(\tau)=-\frac{1}{8N\tau}\frac{\del}{\del\tau}\Biggl(\frac{1}{8N\tau}
\frac{\del g_{\text{conn}}(\tau)}{\del\tau}\Biggr)=-\frac{\del^2 g_{\text{conn}}(\tau)}{\del a^2},\qquad
(a=4N\tau^2).
\end{aligned} 
\label{eq:disc-conn}
\end{equation}
We emphasize that this relation holds exactly at finite $N$.

In the large $N$ scaling limit \eqref{eq:tau},
$\del_\tau g_{\text{conn}}(\tau)$
behaves differently below and above the critical time $\tau=1$ \cite{Okuyama:2018yep}
\begin{equation}
\begin{aligned}
 \del_\tau g_{\text{conn}}(\tau)=\left\{
\begin{aligned}
 &\frac{4N}{\pi}\rt{1-\tau^2},\quad &(\tau<1),\\
&0,\quad &(\tau>1).
\end{aligned}\right.
\end{aligned} 
\label{eq:semi-tau}
\end{equation}
Namely, when $\tau<1$ $\del_\tau g_{\text{conn}}(\tau)$ obeys the semi-circle
law as a function of time \cite{Okuyama:2018yep}, while 
$\del_\tau g_{\text{conn}}(\tau)$ vanishes beyond $\tau=1$.
The regime $\tau<1$ corresponds to the ramp and 
the regime $\tau>1$ corresponds to the plateau.
In the rest of this paper, we will consider the behavior of the SFF
in the plateau regime $\tau>1$.

The vanishing of $\del_\tau g_{\text{conn}}(\tau)$ in the plateau regime $\tau>1$ in 
\eqref{eq:semi-tau} 
is a result of the strict large $N$ limit. At finite $N$, $\del_\tau g_{\text{conn}}(\tau)$
is non-zero but exponentially suppressed when $N\gg1$.
It turns out that this exponentially small correction is non-perturbative
in the $1/N$ expansion, and the study of such non-perturbative corrections
is the main purpose of this paper.
Using the relations \eqref{eq:disc-z^2} and \eqref{eq:disc-conn}, 
the non-perturbative correction to $\del_\tau g_{\text{conn}}(\tau)$
can be obtained once we know the 
non-perturbative correction to the one-point function $\mathcal{Z}(\tau)$.
As we will see shortly, $\mathcal{Z}(\tau)$ in the plateau regime behaves as
\begin{equation}
\begin{aligned}
 \mathcal{Z}(\tau)\sim e^{-NS_{\text{inst}}(\tau)},
\end{aligned} 
\end{equation}
with the ``instanton action'' $S_{\text{inst}}(\tau)$ being
\begin{equation}
\begin{aligned}
 S_{\text{inst}}(\tau)
=2\Bigl[\tau\rt{\tau^2-1}-\text{arccosh}(\tau)\Bigr].
\end{aligned} 
\label{eq:sinst}
\end{equation}
From the relations \eqref{eq:disc-z^2} and \eqref{eq:disc-conn} it follows
that $\del_\tau g_{\text{conn}}(\tau)$
receives the two-instanton correction
\begin{equation}
\begin{aligned}
\del_\tau g_{\text{conn}}(\tau)\sim e^{-2NS_{\text{inst}}(\tau)}.
\end{aligned} 
\label{eq:gconn-sim}
\end{equation}

Let us consider the large $N$ behavior of $\mathcal{Z}(\tau)$ in \eqref{eq:Zexact}
in the plateau regime. Interestingly, the asymptotic behavior
of the Laguerre polynomial in \eqref{eq:Zexact} has been studied in 
a different context in \cite{Gross:1994mr};
exactly the same expression as \eqref{eq:Zexact} appeared in the
study of the partition function of 2d Yang-Mills theory on $S^2$ 
with the identification
\begin{equation}
\begin{aligned}
 \tau~\leftrightarrow~\frac{\pi^2}{A},
\end{aligned} 
\label{eq:dictionary}
\end{equation}
where $A$ is the 't Hooft coupling of 2d Yang-Mills theory.  With this identification,
the plateau regime $\tau>1$ corresponds to the weak coupling phase $A<\pi^2$
of 2d Yang-Mills theory and the transition at $\tau=1$ from the ramp to the plateau 
corresponds to the Douglas-Kazakov phase transition 
of 2d Yang-Mills theory on $S^2$ \cite{Douglas:1993iia}.
The ``instanton action'' $\ga(A/\pi^2)$ 
in the 2d Yang-Mills theory has been computed 
in \cite{Gross:1994mr} using the integral representation of
the Laguerre polynomial and it can be translated to
the instanton action $S_{\text{inst}}(\tau)$ in \eqref{eq:sinst}\footnote{
Interestingly, a similar expression
has also appeared in the instanton computation of the
Gross-Witten-Wadia unitary matrix model (see e.g. eq.(4.23) in \cite{Marino:2008ya}).}
via the dictionary \eqref{eq:dictionary}.

\begin{figure}[thb]
\centering
\includegraphics[width=10cm]{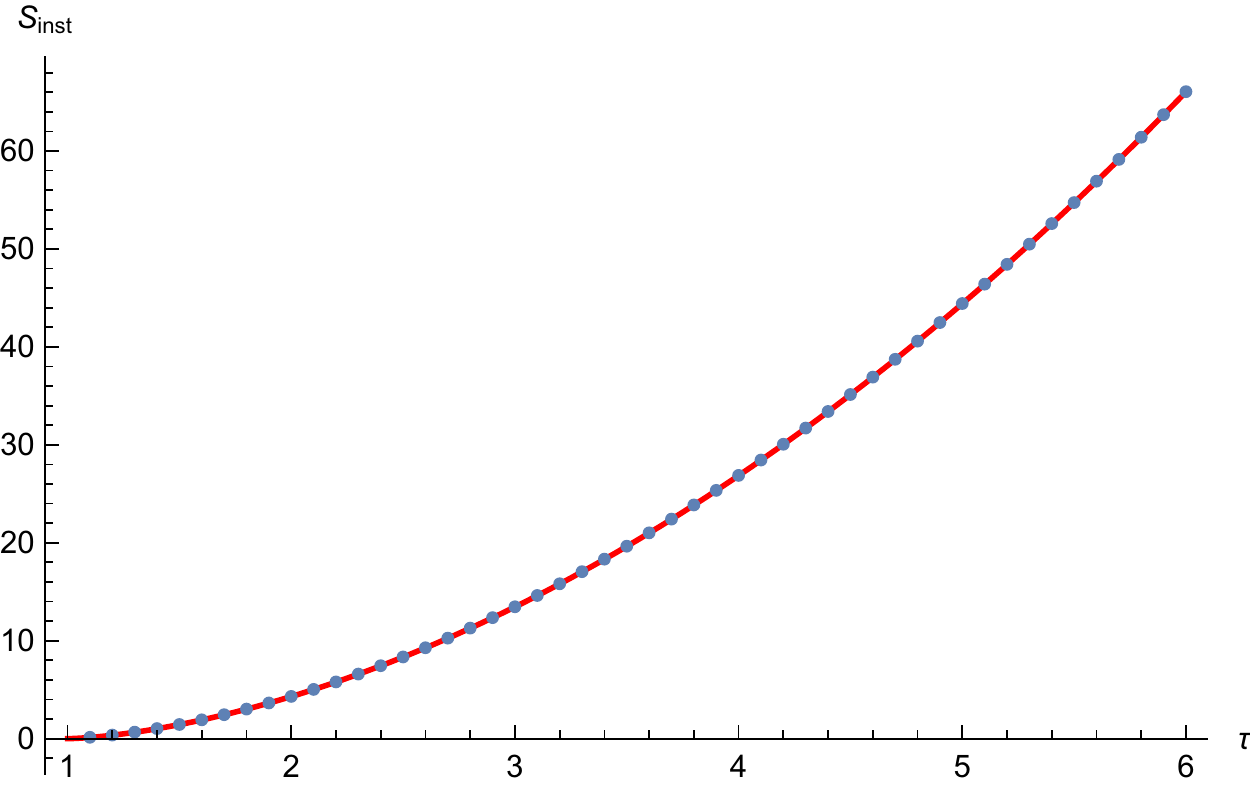}
\caption{Plot of the instanton action $S_{\text{inst}}(\tau)$.
The dots are the exact values of the right hand side of \eqref{eq:sinst-app0}
for $N=500$, while the red curve represents
the analytic form of $S_{\text{inst}}(\tau)$
in \eqref{eq:sinst}.
}
  \label{fig:sinst-bt0}
\end{figure}

We can numerically check the expected behavior of $\del_\tau g_{\text{conn}}(\tau)$
in \eqref{eq:gconn-sim} using the exact form of $\del_\tau g_{\text{conn}}(\tau)$
in \eqref{eq:gconn-exact}. One can extract the instanton action 
from the exact result of $\del_\tau g_{\text{conn}}(\tau)$
in \eqref{eq:gconn-exact} as
\begin{equation}
\begin{aligned}
 S_{\text{inst}}(\tau)\approx -\frac{1}{2N}\log\Big(\del_\tau g_{\text{conn}}(\tau)\Big),\quad
(N\gg1).
\end{aligned} 
\label{eq:sinst-app0}
\end{equation}
In Figure \ref{fig:sinst-bt0}, we plot the exact value of the right hand side of \eqref{eq:sinst-app0}
for $N=500$ in the plateau regime $\tau>1$. One can see from Figure \ref{fig:sinst-bt0}
that the exact result of $\del_\tau g_{\text{conn}}(\tau)$
in \eqref{eq:gconn-exact} correctly reproduces the analytic form
of the instanton action in \eqref{eq:sinst} as expected.

Note that the instanton action $S_{\text{inst}}(\tau)$
in \eqref{eq:sinst} vanishes at the critical time $\tau=1$.
In the context of 2d Yang-Mills theory on $S^2$,
this vanishing behavior of instanton action
is interpreted in \cite{Gross:1994mr}
that the Douglas-Kazakov phase transition is induced by the instantons.
In the case of the SFF, 
it is tempting to speculate that the ramp-plateau transition
is also induced by the instantons.
It would be extremely interesting to
study this picture in the
holographic dual of the SYK model, called JT string in \cite{Saad:2018bqo},
and see how the D-branes in the JT string 
are related to the ramp-plateau transition.

We also notice that the expression of $\mathcal{Z}(\tau)$
in \eqref{eq:Zexact} has another interesting interpretation
in the 4d $\mathcal{N}=4$ super Yang-Mills (SYM) theory. 
The expectation value of 1/2 BPS Wilson loop $W_k$
in 4d $\mathcal{N}=4$ SYM theory with winding number $k$ is given by
\cite{Drukker:2000rr,Drukker:2005kx} \footnote{
Note that the expectation value of the winding Wilson loop in 2d Yang-Mills theory
also has the same expression as $\mathcal{Z}(\tau)$ in \eqref{eq:Zexact} \cite{Gross:1994ub}.}
\begin{equation}
\begin{aligned}
 \bigl\bra W_k\bigr\ket_{\mathcal{N}=4~\text{SYM}}=\frac{1}{N}e^{2N\ka^2}
L_{N-1}^1(-4N\ka^2),
\end{aligned} 
\label{eq:D3}
\end{equation}
where $\ka$ is given by
\begin{equation}
\begin{aligned}
 \ka=\frac{k\rt{\la}}{4N},
\end{aligned} 
\end{equation}
and $\la$ denotes the 't Hooft coupling of  $\mathcal{N}=4$ SYM.
As discussed in \cite{Drukker:2005kx}, in the limit of large winding number $k\sim \mathcal{O}(N)$
\begin{equation}
\begin{aligned}
 k\to\infty,~N\to \infty,\quad \text{with}~~\ka~~\text{fixed},
\end{aligned} 
\end{equation}
the holographically dual object of the winding Wilson loop becomes a D3-brane
embedded in $AdS_5\times S^5$ and $\ka$ corresponds to the electric flux on the D3-brane worldvolume. 
One can see that the Wilson loop in $\mathcal{N}=4$ SYM in \eqref{eq:D3} and  $\mathcal{Z}(\tau)$ in \eqref{eq:Zexact} 
are related by an analytic continuation 
\begin{equation}
\begin{aligned}
 \ka~\leftrightarrow~\ri\tau.
\end{aligned} 
\label{eq:ka-tau}
\end{equation}
In the computation of D3-brane action in \cite{Drukker:2005kx},
the metric on the D3-brane is taken to be Euclidean. After the above analytic continuation
$\ka\to\ri\tau$, $\mathcal{Z}(\tau)$ formally corresponds to
a Lorentzian signature D3-brane with an electric field $\tau$. 
In this picture, the critical value $\tau=1$ can be thought of as the
critical electric field for the threshold of 
holographic Schwinger effect \cite{Semenoff:2011ng}.
Appearance of the Lorentzian signature is not surprising
since the SFF captures the chaotic behavior
of a system in the real time dynamics.

\section{Eigenvalue instanton in the spectral form factor \label{sec:inst}}
The large $N$ behavior of $\mathcal{Z}(\tau)$ in \eqref{eq:Zexact}
has been already studied in \cite{Gross:1994mr}
in the context of 2d Yang-Mills theory on $S^2$,
where the scaling limit \eqref{eq:tau}
corresponds to the 't Hooft limit in the 2d Yang-Mills theory.
Here we repeat the same analysis from the viewpoint of eigenvalue instanton
in the Gaussian matrix model,
in a similar manner as in \cite{Kawamoto:2008gp,Hartnoll:2006is}
for the large winding number limit of
1/2 BPS Wilson loop in 4d $\mathcal{N}=4$ SYM.

Let us consider the eigenvalue integral representation of the
one-point function
$\mathcal{Z}(\tau)$ defined in \eqref{eq:Zdef}
\begin{equation}
\begin{aligned}
 \mathcal{Z}(\tau)
&=\int \prod_{i=1}^N d\mu_i e^{-\frac{N}{2}\mu_i^2} \prod_{1\leq i<j\leq N}(\mu_i-\mu_j)^2 \sum_{i=1}^N e^{2\ri N\tau \mu_i}.
\end{aligned} 
\end{equation}
Here we have ignored the overall normalization constant for simplicity.
Integrating over the $N-1$ eigenvalues 
which do not couple to the operator insertion
$e^{2\ri N  \tau \mu_i}$,
$\mathcal{Z}(\tau)$ can be written as an integral for a single eigenvalue $x$
which has a non-zero coupling to the operator insertion $e^{2\ri N\tau x}$
\begin{equation}
\begin{aligned}
 \mathcal{Z}(\tau)=\int dx \,e^{-NV_{\text{eff}}(x)+2\ri N\tau x},
\end{aligned} 
\label{eq:x-int}
\end{equation}
where the effective potential $V_{\text{eff}}(x)$ for the eigenvalue $x$ is defined by
integrating out the other $N-1$ eigenvalues
\begin{equation}
\begin{aligned}
 e^{-NV_{\text{eff}}(x)}=e^{-\frac{N}{2}x^2}
\int \prod_{i=1}^{N-1} d\mu_i e^{-\frac{N}{2}\mu_i^2}(x-\mu_i)^2
\prod_{1\leq i<j\leq N-1}(\mu_i-\mu_j)^2 .
\end{aligned} 
\end{equation}
In the scaling limit \eqref{eq:tau},
the $x$-integral \eqref{eq:x-int}
can be evaluated by the saddle point approximation.
The saddle point $x=x_*$ is determined by the condition
\begin{equation}
\begin{aligned}
 V'_{\text{eff}}(x_*)=2\ri \tau,
\end{aligned} 
\label{eq:saddle-eq}
\end{equation}
where the prime in $V'_{\text{eff}}(x)$ denotes 
the derivative
with respect to $x$.
In the large $N$ limit $V'_{\text{eff}}(x)$ is 
written as a sum of the contributions
from the Gaussian potential $x^2/2$ and the Coulomb repulsion from the
other $N-1$ eigenvalues
\begin{equation}
\begin{aligned}
V'_{\text{eff}}(x)=x-2\int_{-2}^2 dy\frac{\rho(y)}{x-y}.
\end{aligned} 
\label{eq:dVeff}
\end{equation}
Here  $\rho(y)$ denotes the eigenvalue density
given by the Wigner semi-circle law
\begin{equation}
\begin{aligned}
 \rho(y)=\frac{1}{2\pi}\rt{4-y^2},
\end{aligned} 
\end{equation}
and the $N-1$ eigenvalues are distributed along the cut $[-2,2]$ in the large $N$
limit.
The integral in \eqref{eq:dVeff} is easily evaluated as
\begin{equation}
\begin{aligned}
V'_{\text{eff}}(x)=\rt{x^2-4}. 
\end{aligned} 
\end{equation}
Using this expression, we find a non-trivial solution for the
saddle point equation \eqref{eq:saddle-eq}
\begin{equation}
\begin{aligned}
 x_*=2\ri v_*,\quad v_*=\rt{\tau^2-1}.
\end{aligned} 
\label{eq:vsaddle}
\end{equation}
Namely, in the scaling limit 
\eqref{eq:tau} one eigenvalue $x$ is pulled out from the cut $[-2,2]$
to the saddle point $x_*=2\ri v_*$ on the imaginary axis.
This is interpreted as the eigenvalue instanton \cite{David:1990sk,David:1992za}
and
the instanton action is given by
\begin{equation}
\begin{aligned}
 S_{\text{inst}}(\tau)=\Bigl[V_{\text{eff}}(x_*)-2\ri\tau x_*\Bigr]
-V_{\text{eff}}(x_0) ,
\end{aligned} 
\end{equation}
where $x_0$ is a reference point on the cut $x_0\in[-2,2]$.
Since $V_{\text{eff}}(x)$ is constant along the cut, one can choose $x_0=0$ without loss of generality.
Then the instanton action is written as a change of effective potential
as we move the eigenvalue along the imaginary axis from $x=0$ to $x=2\ri v_*$
\begin{equation}
\begin{aligned}
 S_{\text{inst}}(\tau)=\int_0^{x_*}dx \Bigl[V'_{\text{eff}}(x)-2\ri\tau\Bigr]
=4\int_0^{v_*}dv\Bigl[-\rt{1+v^2}+\tau\Bigr].
\end{aligned} 
\label{eq:sinst-vint}
\end{equation}
In the last equality we have changed the integration variable $x=2\ri v$.
One can easily show that \eqref{eq:sinst-vint}
reproduces the instanton action \eqref{eq:sinst}
obtained in \cite{Gross:1994mr}.

Note that in the case of large winding Wilson loop in $\mathcal{N}=4$ SYM
one eigenvalue is pulled out of the cut $[-2,2]$ to the saddle point on the real axis
\cite{Kawamoto:2008gp,Hartnoll:2006is}, while
in our case the saddle point is on the imaginary axis.

We can repeat the same saddle point analysis for the
SFF defined by the matrix integral \eqref{eq:gdef}.
In the scaling limit \eqref{eq:tau}, $g_{\text{conn}}(\tau)$
is written as the integral of two eigenvalues $x,y$
\begin{equation}
\begin{aligned}
 g_{\text{conn}}(\tau)\sim \int dxdy (x-y)^2e^{-NV_{\text{eff}}(x)-NV_{\text{eff}}(y)
+2\ri N\tau(x-y)}.
\end{aligned} 
\label{eq:gconn-xy}
\end{equation}
At the leading order in the large $N$ expansion, the interaction between the two 
eigenvalues $(x-y)^2$ can be neglected and
the saddle point equation becomes
\begin{equation}
\begin{aligned}
 V'_{\text{eff}}(x_*)=2\ri\tau,\quad V'_{\text{eff}}(y_*)=-2\ri\tau.
\end{aligned} 
\end{equation}
The solution of this saddle point equation
is given by a complex conjugate pair
of points on the imaginary axis
\begin{equation}
\begin{aligned}
 x_*=2\ri v_*,\quad y_*=\b{x}_*=-2\ri v_*,
\end{aligned} 
\label{eq:saddle-xy}
\end{equation}
where $v_*$ is defined in \eqref{eq:vsaddle}.
By evaluating the saddle point value of the effective potential
one can see that $g_{\text{conn}}(\tau)$ behaves as
the two-instanton correction
\begin{equation}
\begin{aligned}
 g_{\text{conn}}(\tau)\sim e^{-2NS_{\text{inst}}(\tau)}.
\end{aligned} 
\end{equation}
This is consistent with the behavior \eqref{eq:gconn-sim} expected from 
the exact finite $N$ relations \eqref{eq:disc-z^2} and \eqref{eq:disc-conn}.

Before closing this section, we note that the $\tau$-derivative of instanton action 
in \eqref{eq:sinst-vint} has a simple form. The $\tau$-derivative of the boundary term
$v=v_*$ in \eqref{eq:sinst-vint} 
vanishes since $\rt{1+v_*^2}=\tau$ and we find
\begin{equation}
\begin{aligned}
 \del_\tau S_{\text{inst}}(\tau)=4\int_0^{v_*}dv=4v_*=4\rt{\tau^2-1}.
\end{aligned} 
\label{eq:dSinst}
\end{equation}
Of course, this agrees with the $\tau$-derivative of $S_{\text{inst}}(\tau)$
in \eqref{eq:sinst}.
\section{$1/N$ expansion around the instanton \label{sec:1/N}}

In this section we compute the $1/N$ corrections to
$\del_\tau g_{\text{conn}}(\tau)$ 
around the instanton 
in \eqref{eq:gconn-sim}.
This is obtained once we know the $1/N$ corrections of $\mathcal{Z}(\tau)$
using the relations \eqref{eq:disc-z^2} and \eqref{eq:disc-conn}.
This problem was studied in \cite{Gross:1994mr} using
the integral representation of Laguerre polynomial.
Here we follow the approach in \cite{Drukker:2005kx}
using the differential equation satisfied by $\mathcal{Z}(\tau)$.
From the differential equation for the Laguerre polynomial
\begin{equation}
\begin{aligned}
 \left[x\frac{d^2}{ dx^2}+(\al+1-x)\frac{d}{ dx}+n\right]L^\al_n(x)=0,
\end{aligned} 
\end{equation}
one can show that $\mathcal{Z}(\tau)$ obeys
\begin{equation}
\begin{aligned}
\Bigl[\tau\del_\tau^2+3\del_\tau+16N^2\tau(1-\tau^2)\Bigr] \mathcal{Z}(\tau)=0.
\end{aligned} 
\end{equation}
As discussed in
\cite{Drukker:2000rr},
it is convenient to recast this into the equation for the  ``free energy'' 
$\mathcal{F}(\tau)=-\frac{1}{N}\log \mathcal{Z}(\tau)$
\begin{equation}
\begin{aligned}
 (\del_\tau \mathcal{F})^2-\frac{1}{N\tau}(\tau \del_\tau^2 \mathcal{F}+3\del_\tau \mathcal{F})+16(1-\tau^2)=0.
\end{aligned} 
\label{eq:DF-eq}
\end{equation}
One can solve this equation order by order in the $1/N$ expansion
\begin{equation}
\begin{aligned}
 \mathcal{F}(\tau)=\sum_{\ell=0}^\infty \frac{1}{N^\ell}\mathcal{F}_\ell(\tau).
\end{aligned} 
\label{eq:Feq}
\end{equation}
At the leading order we have
\begin{equation}
\begin{aligned}
 \del_\tau \mathcal{F}_0(\tau)=4\rt{\tau^2-1},
\end{aligned} 
\end{equation}
and from \eqref{eq:dSinst} the leading term is given by the instanton action
in \eqref{eq:sinst},
as expected
\begin{equation}
\begin{aligned}
 \mathcal{F}_0(\tau)=S_{\text{inst}}(\tau).
\end{aligned} 
\end{equation}
One can systematically find the higher order corrections $\mathcal{F}_\ell(\tau)$
by solving the differential equation \eqref{eq:DF-eq} order by order in $1/N$.
In this way we find the $1/N$ expansion of $\mathcal{Z}(\tau)=e^{-N\mathcal{F}(\tau)}$
\begin{equation}
\begin{aligned}
 \mathcal{Z}(\tau)=(-1)^{N-1}\frac{e^{-NS_\text{inst}(\tau)}}{\rt{32\pi N}\tau^{\frac{3}{2}}(\tau^2-1)^{\qu}}
\sum_{\ell=0}^\infty
\frac{1}{N^\ell}z_\ell(\tau),
\end{aligned}
\label{eq:Zexpand} 
\end{equation}
where the first few terms read
\begin{equation}
\begin{aligned}
z_0(\tau)&=1,\\
z_1(\tau)&=-
\frac{8\tau^4-12\tau^2+9}{96\tau(\tau^2-1)^{3/2}},\\
z_2(\tau)&=\frac{64 \tau ^8-192 \tau ^6+288 \tau ^4+360 \tau
   ^2-135}{18432 \tau ^2 \left(\tau ^2-1\right)^3},\\
z_3(\tau)&=\frac{71168 \tau ^{12}-320256 \tau ^{10}+554688 \tau
   ^8-518400 \tau ^6-340200 \tau ^4+170100 \tau
   ^2-42525}{26542080 \tau ^3 \left(\tau
   ^2-1\right)^{9/2}}.
\end{aligned} 
\end{equation}
Some comments on \eqref{eq:Zexpand} are in order here:
\begin{itemize}
 \item[(i)] The overall factor $1/\rt{32\pi N}$ in \eqref{eq:Zexpand} cannot be determined 
by the differential equation \eqref{eq:DF-eq} alone.
This factor can be obtained by a careful analysis of the Gaussian integral
around the saddle point \cite{Gross:1994mr}.
  \item[(ii)] As emphasized in \cite{Gross:1994mr}, 
the large $N$ limit of $\mathcal{Z}(\tau)$ in \eqref{eq:Zexpand}
has an alternating sign $(-1)^{N-1}$, i.e.
$\mathcal{Z}(\tau)$ is negative for even $N$ and positive
for odd $N$.
However, this sign is absent in the 
disconnected part of the
SFF $g_{\text{disc}}(\tau)=\mathcal{Z}(\tau)^2$.
The connected part $g_{\text{conn}}(\tau)$ does not have this sign
either because of 
the relation \eqref{eq:disc-conn}. \footnote{
In the partition function of 2d Yang-Mills theory on $S^2$,
this sign is canceled in the final result after a careful
analysis of the extra sign coming from the modular transformation
of the partition function \cite{Okuyama:2018kic}.}
\end{itemize} 

Once we know the $1/N$ expansion of $\mathcal{Z}(\tau)$ in \eqref{eq:Zexpand},
we can easily find the $1/N$ expansion of $\del_\tau g_{\text{conn}}(\tau)$
using the relations \eqref{eq:disc-z^2} and \eqref{eq:disc-conn}:
\begin{equation}
\begin{aligned}
 \del_\tau g_{\text{conn}}(\tau)=\frac{e^{-2NS_{\text{inst}}(\tau)}}{4\pi\tau(\tau^2-1)}
\sum_{\ell=0}^\infty
\frac{1}{N^\ell} g_\ell(\tau),
\end{aligned} 
\label{eq:gconn-sub}
\end{equation}
where the first few terms are given by
\begin{equation}
\begin{aligned}
g_0(\tau)&=1,\\
g_1(\tau)&=-\frac{8 \tau ^4+12 \tau ^2-3}{48 \tau  \left(\tau
   ^2-1\right)^{3/2}},
\\
g_2(\tau)&=\frac{64 \tau ^8+192 \tau ^6+1248 \tau ^4-360 \tau
   ^2+81}{4608 \tau ^2 \left(\tau ^2-1\right)^3},\\
g_3(\tau)&=\frac{15872 \tau ^{12}-94464 \tau ^{10}-7488 \tau
   ^8-1054080 \tau ^6+249480 \tau ^4-127980 \tau
   ^2+23085}{3317760 \tau ^3 \left(\tau
   ^2-1\right)^{9/2}}.
\end{aligned} 
\end{equation}
We have checked numerically that this expansion is consistent with the exact result
of $\del_\tau g_{\text{conn}}(\tau)$  in \eqref{eq:gconn-exact}.

Apparently, the $1/N$ expansion of $\del_\tau g_{\text{conn}}(\tau)$ in \eqref{eq:gconn-sub}
breaks down at $\tau=1$. However, this is just an artifact of the expansion  \eqref{eq:gconn-sub}
and the exact result of $\del_\tau g_{\text{conn}}(\tau)$ in \eqref{eq:gconn-exact}
is perfectly smooth at $\tau=1$.
A similar phenomenon was observed in the Gross-Witten-Wadia model in \cite{Ahmed:2017lhl}.
As discussed in \cite{Ahmed:2017lhl},
we can improve the expansion near the critical point using the analogue of the
uniform WKB expansion (see \cite{nist} for a uniform expansion of the Laguerre
polynomial in terms of the Airy function). It would be interesting to
study the detail of this uniform expansion.

\section{Comment on the $\bt\ne0$ case \label{sec:beta}}
In this section, we will briefly comment on the case of non-zero $\bt$.
The SFF with non-zero $\bt$ is defined by
\begin{equation}
\begin{aligned}
 g(\bt,\tau)=\Bigl\bra \Tr e^{-(\bt+2N\ri\tau)H}\Tr e^{-(\bt-2N\ri\tau)H}\Bigr\ket
.
\end{aligned} 
\end{equation}
Again, it is decomposed into the disconnected part 
$g_{\text{disc}}(\bt,\tau)$ and the connected part $g_{\text{conn}}(\bt,\tau)$.
The exact form of $\del_\tau g_{\text{conn}}(\bt,\tau)$
was obtained in \cite{Okuyama:2018yep}
\begin{equation}
\begin{aligned}
\del_\tau g_{\text{conn}}(\bt,\tau)
=\frac{2N^2}{\beta}e^{-4N\tau^2+\frac{\bt^2}{N}}
\text{Im}\left[L_N\Bigl(4Nw^2\Bigr)
L_{N-1}\Bigl(4N\b{w}^2\Bigr)\right],
\end{aligned} 
\end{equation}
with $w,\b{w}$ being
\begin{equation}
\begin{aligned}
 w=\tau-\ri\frac{\bt}{2N},\quad
\b{w}=\tau+\ri\frac{\bt}{2N}.
\end{aligned} 
\end{equation}
When $\bt\ne0$, we do not have a simple relation between 
$g_{\text{disc}}$ and $g_{\text{conn}}$ as in \eqref{eq:disc-conn},
and hence the information of the $1/N$ expansion of the one-point function alone
is not enough to find the  $1/N$ expansion of $\del_\tau g_{\text{conn}}(\bt,\tau)$.

\begin{figure}[thb]
\centering
\includegraphics[width=10cm]{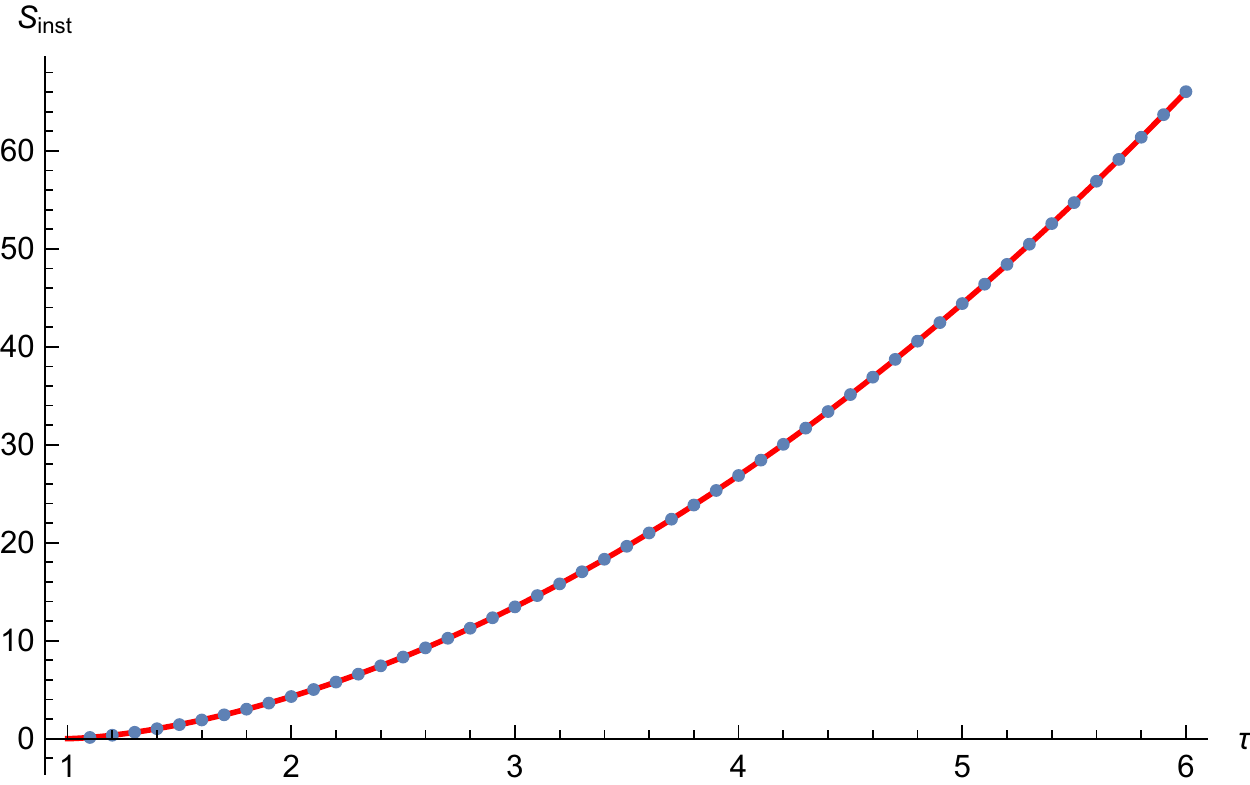}
\caption{Plot of the instanton action $S_{\text{inst}}(\bt,\tau)$ for $\bt=5$.  
 The dots are the exact values of the right hand side of \eqref{eq:sinst-app}
for $\bt=5, N=500$, while the red curve represents
the analytic form of $S_{\text{inst}}(\tau)$
at $\bt=0$ in \eqref{eq:sinst}.
}
  \label{fig:sinst-bt5}
\end{figure}

We expect that in the scaling limit \eqref{eq:tau} with finite $\bt\sim \mathcal{O}(N^0)$,
$\del_\tau g_{\text{conn}}(\bt,\tau)$ is exponentially small in the plateau regime
$\tau>1$
\begin{equation}
\begin{aligned}
 \del_\tau g_{\text{conn}}(\bt,\tau)\sim e^{-2NS_{\text{inst}}(\bt,\tau)}.
\end{aligned} 
\end{equation}
In a similar manner as \eqref{eq:sinst-app0},
we can numerically extract the instanton action $S_{\text{inst}}(\bt,\tau)$
from the exact result of $\del_\tau g_{\text{conn}}(\bt,\tau)$ at finite $N$
\begin{equation}
\begin{aligned}
 S_{\text{inst}}(\bt,\tau)\approx -\frac{1}{2N}\log \Bigl[\del_\tau g_{\text{conn}}(\bt,\tau)\Bigr],
\qquad (N\gg1).
\end{aligned} 
\label{eq:sinst-app}
\end{equation}
We observed numerically that $S_{\text{inst}}(\bt,\tau)$ is independent of $\bt$
\begin{equation}
\begin{aligned}
 S_{\text{inst}}(\bt,\tau)=S_{\text{inst}}(\tau),
\end{aligned} 
\end{equation}
where the instanton action $S_{\text{inst}}(\tau)$ is given by \eqref{eq:sinst}.
See Figure
\ref{fig:sinst-bt5} for the plot of \eqref{eq:sinst-app} at $\bt=5$
with $N=500$.

This $\bt$-independence of the instanton action
can be understood from the eigenvalue integral considered in \eqref{eq:gconn-xy}.
When $\bt\ne0$, \eqref{eq:gconn-xy} is modified to
\begin{equation}
\begin{aligned}
 g_{\text{conn}}(\bt,\tau)\sim \int dxdy (x-y)^2e^{-NV_{\text{eff}}(x)-
NV_{\text{eff}}(y)+2\ri N\tau(x-y)+\bt(x+y)}.
\end{aligned} 
\end{equation}
When evaluating this integral in the saddle point approximation,
the location of saddle point would be slightly shifted from the $\bt=0$ case in \eqref{eq:saddle-xy}.
However, this shift is subleading in the $1/N$ expansion as long as 
$\bt$ is order $N^0$ and hence the value of the effective potential at the saddle point 
does not change at the leading order in the large $N$ expansion.
This explains the $\bt$-independence of the instanton action
observed numerically in Figure \ref{fig:sinst-bt5}.

We expect that the $1/N$ expansion around the instanton for $\bt=0$ in \eqref{eq:gconn-sub}
is modified when $\bt\ne0$ and the $\bt$-dependence arises at the subleading order
in the $1/N$ expansion
\begin{equation}
\begin{aligned}
  \del_\tau g_{\text{conn}}(\bt,\tau)=\frac{e^{-2NS_{\text{inst}}(\tau)}}{4\pi\tau(\tau^2-1)}
\sum_{\ell=0}^\infty
\frac{1}{N^\ell} g_\ell(\bt,\tau).
\end{aligned} 
\end{equation}
It would be interesting to study the
$\bt$-dependence of the expansion coefficients $g_\ell(\bt,\tau)$.
We leave this as a future problem.

\section{Conclusion and outlook \label{sec:conc}}
In this paper we have studied the exponentially small correction to the
connected part of the SFF in the GUE random matrix model
in the plateau regime.
We have argued that such a correction comes from the eigenvalue instanton
in the matrix model, where two eigenvalues are pulled out from the cut
to the complex conjugate pair of saddle points \eqref{eq:saddle-xy}
on the imaginary axis. We found the
analytic form of the instanton action \eqref{eq:sinst} as a function of
time and demonstrated that the $1/N$ corrections around the instanton configuration
can be systematically obtained from the differential equation \eqref{eq:DF-eq}
and the non-trivial relation \eqref{eq:disc-conn} between the connected and the disconnected
part of the SFF. 

At the technical level, the non-perturbative corrections in the SFF in the GUE matrix model
is formally equivalent to the same problem in the 2d Yang-Mills theory
on $S^2$. 
In this correspondence, 
the ramp-plateau transition in the SFF
is related to the Douglas-Kazakov phase transition in the 2d Yang-Mills theory on $S^2$.
Also, we have seen that this problem is related to
the large winding number limit of the 1/2 BPS Wilson loop
in 4d $\mathcal{N}=4$ SYM after an analytic continuation
\eqref{eq:ka-tau}.

The study in this paper is strongly motivated by the
recent discussion of a possible interpretation of the
plateau in the SFF of the SYK model \cite{Cotler:2016fpe,Saad:2018bqo}.
It was advocated in \cite{Cotler:2016fpe,Saad:2018bqo}
that the plateau in the SYK model 
is closely related to the non-perturbative effect in the $1/N$ expansion, which is
identified with the effect of D-branes in the holographic dual
string theory of SYK model, dubbed the ``JT string'' in \cite{Saad:2018bqo}.
In this paper, we have seen that this picture is indeed realized in the simple example
of the Gaussian matrix model where the role of D-brane is played by the eigenvalue 
instanton in the matrix model. 
In the analogy with the large winding Wilson loop
in $\mathcal{N}=4$ SYM,
the appearance of a complex conjugate pair of saddle
points can be thought of as a pair creation of D-branes by the Schwinger effect.
It would be very interesting to see how
such non-perturbative effects arise in the case of the SYK model.
Perhaps, one can use the relation between the spectrum of SYK
model and the $q$-Hermite polynomials discussed in \cite{Garcia-Garcia:2017pzl}.

The non-perturbative correction to the SFF
is an interesting problem in its own right,
regardless of the connection to the SYK model and quantum gravity.
In this paper, we have seen that
the ramp-plateau transition of
the SFF of GUE matrix model
can be thought of as a kind
of phase transition induced by the eigenvalue instantons.
It would be interesting to study the non-perturbative correction
in other ensembles, such as GOE and GSE,
and see if the above picture of phase transition applies to
other models as well.
The saddle point analysis for the SFF of GOE and GSE
might be tractable, as we did for GUE in section \ref{sec:inst}.
Also, using the known result of 1/2 BPS Wilson loops in 4d $\mathcal{N}=4$
SYM with gauge group $SO(N)$ and $Sp(N)$ \cite{Fiol:2014fla},
it would be possible to write down the exact SFF of GOE and GSE
at finite $N$ and study its large $N$ limit.
We leave this as an interesting future problem. 

\acknowledgments
I would like to thank Shinji Hirano, Tadakatsu Sakai, Yuki Sato and  Masaki Shigmeori for discussion.  
This work was supported in part by JSPS KAKENHI Grant Number 16K05316.


\end{document}